\documentclass[russian]{article}
\usepackage[T2A,T2A,T2A,T1]{fontenc}
\usepackage[koi8-r]{inputenc}
\usepackage{amsmath}
\usepackage{esint}

\makeatletter

\AtBeginDocument{\DeclareFontEncoding{T2A}{}{}}

\@ifundefined{date}{}{\date{}}
\AtBeginDocument{\DeclareFontEncoding{T2A}{}{}}\AtBeginDocument{\DeclareFontEncoding{T2A}{}{}}\usepackage{babel}

\usepackage{babel}

\makeatother

\usepackage{babel}
\begin{document}

\title{Обобщение дираковского сопряжения в супералгебраической теории спиноров}

\author{В.В.Монахов}

\maketitle
\begin{center}
Санкт-Петербургский государственный университет 
\par\end{center}

\begin{center}
v.v.monahov@spbu.ru 
\par\end{center}

\begin{center}
(Статья принята в журнал <<Теоретическая и математическая физика>>,
2019
\par\end{center}

\begin{center}
Accepted in <<Theoretical and Mathematical Physics>>, 2019) 
\par\end{center}
\begin{abstract}
В супералгебраическом представлении спиноров, использующем грассмановы
плотности и производные по ним, введено обобщение дираковского сопряжения,
обеспечивающее Лоренц-ковариантные преобразования сопряженных спиноров.
Показано, что сигнатура обобщенных гамма-матриц и их количество, а
также разложение вторичного квантования по импульсам задаются вариантом
обобщенного дираковского сопряжения и требованием сохранения при преобразованиях
спиноров и сопряженных спиноров их CAR-алгебры. 
\end{abstract}
Ключевые слова: вторичное квантование, CAR-алгебра, алгебра Клиффорда,
матрицы Дирака, спиноры, дираковское сопряжение, преобразования Лоренца,
Лоренц-ковариантность, причинность, оператор заряда

\section{Введение}

Уравнение Дирака и дираковское сопряжение играют важнейшую роль в
теории спиноров (фермионов). Исторически сложилось несколько подходов
к теориям спиноров {[}1{]}-{[}3{]}.

Первоначально рассматривались так называемые ковариантные спиноры
{[}4{]}, {[}5{]}, открытые Картаном. В дальнейшем основным стал подход,
основанный на алгебрах Клиффорда {[}1{]}-{[}3{]}, {[}6{]}-{[}13{]},
в нем рассматриваются алгебраические спиноры. Это a-спиноры {[}2{]}
\textendash{} идеалы, образованные с помощью умножения всех элементов
клиффордовой алгебры на примитивный идемпотент, и e-спиноры {[}2{]}
\textendash{} идеалы, образованные с помощью умножения элементов клиффордовой
алгебры на произвольный идемпотент. Также рассматриваются так называемые
операторные спиноры {[}2{]}, {[}3{]}, {[}11{]}-{[}13{]}, не эквивалентные
дираковским фермионам, и которым пока не найдено соответствия среди
физических частиц.

Несмотря на доказанную для лоренцевской метрики эквивалентность ковариантных
и алгебраических спиноров {[}2{]}, в случае искривленного пространства
их расслоения ведут себя по-разному {[}2{]}, поэтому пока нет однозначности
в том, какой из этих типов спиноров соответствует физическим частицам
\textendash{} либо же физическим частицам соответствует еще какой-то
тип спиноров, который в плоском пространстве-времени эквивалентен
им.

В обычном подходе {[}1{]}, {[}10{]} при рассмотрении спиноров а качестве
первичной рассматривается алгебра Клиффорда. В ней рассматривается
группа Липшица и показывается, что отображения ее подгрупп (спинорных
групп) на группы псевдоортогональных вращений векторного базиса алгебры
Клиффорда являются двукратными накрывающими. Спиноры являются линейными
представлениями спинорных групп. Пространство спиноров является идеалом,
образованным с помощью умножения всех элементов алгебры Клиффорда
на примитивный эрмитовый идемпотент. Данный идемпотент рассматривается
как клиффордовый вакуум. (В случае сигнатуры пространства-времени
с $(p-q)\,mod\,8=1$ идеал не является минимальным и состоит из прямой
суммы двух идеалов).

В работах {[}14{]}-{[}16{]} автором разработан супералгебраический
подход к построению представлений алгебры Клиффорда с инволюцией (алгебраического
обобщения алгебры гамма-матриц Дирака), операторов поля фермионов,
физического вакуума и преобразований Лоренца, позволяющий взглянуть
на теорию алгебраических спиноров и клиффордовых алгебр с новой точки
зрения. В этом подходе в качестве первичных рассматриваются грассмановы
переменные и производные по ним, и из этих переменных строятся базисные
клиффордовы векторы \textendash{} обобщения гамма-матриц Дирака. Супералгебраическое
представление является фермионным аналогом представления Баргмана-Фока
{[}17{]}, используемого для бозонов, альтернативным предложенному
ранее в {[}18{]}.

В работе {[}16{]} для пространства с сигнатурой $(p,q)=(1,3)$ удалось
построить Лоренц-ковариантную супералгебраическую теорию спиноров
и дираковски сопряженных к ним, удовлетворяющих уравнению Дирака и
соответствующих теории вторичного квантования. В том числе удалось
построить Лоренц-инвариантный оператор физического вакуума. Предложенный
подход соответствует построению физических теорий вторичного квантования
на основе супералгебр с использованием грассмановых переменных {[}19{]}-{[}22{]},
и отличается тем, что используются плотности грассмановых переменных
в импульсном пространстве и производные по ним, а гамма-матрицы Дирака
строятся из этих плотностей и производных.

В данной работе предложенный подход развивается и обобщается на пространства
произвольной размерности и сигнатуры. 

Статья состоит из двух частей. Первая часть (разделы 2-4) посвящена
выводу общего вида дираковски сопряженных супералгебраических спиноров,
обеспечивающего их Лоренц-ковариантность.

В разделе <<2 Грассмановы плотности, эрмитово сопряжение, скалярное
произведение и операторы>> вводятся основные понятия, следующие из
работы {[}16{]}, используемые далее. Основа подхода \textendash{}
то, что базисные клиффордовы векторы рассматриваются как составные
объекты, построенные из грассмановых плотностей и производных по ним. 

В разделе <<3 Обобщенное дираковское сопряжение операторов для сигнатуры
пространства-времени (1, n-1)>> в рамках предлагаемого формализма
обобщается понятие дираковского сопряжения и выводится общий вид дираковски
сопряженных супералгебраических спиноров в n-мерном псевдоевклидовом
пространстве с одной времениподобной осью.

В разделе <<4 Обобщенное дираковское сопряжение для произвольной
сигнатуры пространства-времени>> полученный результат обобщается
на псевдоевклидовы пространства произвольной сигнатуры.

Вторая часть статьи (разделы 5-7) посвящена изучению следствий из
первой части в частном случае четырехмерного пространства-времени.

Раздел <<5 Преобразования, порожденные четырьмя грассмановыми плотностями
и производными по ним>> посвящен выводу общего вида преобразований,
сохраняющих CAR-алгебру супералгебраических спиноров. Показано существование
двух дополнительных аналогов гамма-матриц $\hat{\gamma}^{6}$ и $\hat{\gamma}^{7}$
по сравнению с обычной теорией Дирака (что связано с возможностью
в супералгебраическом подходе перемешивания спиноров и дираковски
сопряженных спиноров), а также наличие дополнительных преобразований
супералгебраических спиноров помимо лоренцевских вращений.

В разделе <<6 Разложение оператора поля по импульсам для пространства
с одной времениподобной осью>> показано, что первая часть этих дополнительных
преобразований ведет к появлению разложения вторичного квантования
решений уравнения Дирака.

В разделе <<7 Оператор заряда для пространства с одной времениподобной
осью>> показано, что вторая часть дополнительных преобразований,
связанная с вращениями в плоскости базисных векторов $\hat{\gamma}^{6}$
и $\hat{\gamma}^{7}$, ведет к существованию оператора заряда, имеющего
аналог в теории вторичного квантования и обычно отождествляемого с
оператором электрического заряда.

\section{Грассмановы плотности, эрмитово сопряжение, скалярное произведение
и операторы}

Рассмотрим грассмановы плотности $\theta^{a}(p)$ и производные по
ним $\partial/\partial\theta^{a}(p)$ {[}16{]}, обладающие антикоммутационными
соотношениями 
\begin{equation}
\{\frac{\partial}{\partial\theta^{b}(p)},\theta^{a}(p')\}=\delta_{b}^{a}\delta(p-p'),
\end{equation}
\begin{equation}
\{\frac{\partial}{\partial\theta^{a}(p)},\frac{\partial}{\partial\theta^{b}(p')}\}=\{\theta^{a}(p),\theta^{b}(p')\}=0.
\end{equation}
где $p$ \textendash{} непрерывный параметр (вообще говоря, многомерный),
$\delta(p-p')$ \textendash{} дельта-функция Дирака (вообще говоря,
многомерная), $\delta_{b}^{a}$ \textendash{} дельта-символ Кронекера,
а $a$ и $b$ \textendash{} индексы, меняющиеся в пределах от 1 до
$n_{G}$, где $n_{G}$ \textendash{} число независимых грассмановых
переменных при заданном значении $p$.

Формулы (1), (2) называют каноническими антикоммутационными соотношениями,
они порождают CAR-алгебру.

Эрмитово сопряженной к величине $\theta^{a}(p)$ является производная
$\theta^{a}(p)^{+}=\partial/\partial\theta^{a}(p)$, а к величине
$k\theta^{a}(p)$ величина $k^{*}\partial/\partial\theta^{a}(p)$,
где $k$ \textendash{} числовой коэффициент, а $k^{*}$ \textendash{}
комплексно сопряженный к нему.

Пусть величины $\Psi$ и $\Phi$ являются линейной комбинацией образующих
для заданных значений $p$ и $p'$: 
\begin{align*}
\Psi & =\psi^{\alpha}(p)\frac{\partial}{\partial\theta^{\alpha}(p)}+\psi_{\tau}(p)\theta^{\tau}(p),\\
\Phi & =\phi^{\beta}(p')\frac{\partial}{\partial\theta^{\beta}(p')}+\phi_{\rho}(p')\theta^{\rho}(p'),
\end{align*}
где $\psi^{\alpha}(p),\psi_{\tau}(p),\phi^{\beta}(p'),\phi_{\rho}(p')$
комплексные коэффициенты.

Из (1) и (2) следует

\begin{align*}
\{\Psi^{+},\Phi\} & =(\psi^{\alpha}(p)^{*}\phi^{\alpha}(p)+\psi_{\tau}(p)^{*}\phi_{\tau}(p)\,)\delta(p-p'),
\end{align*}
поэтому результат является комплексным числом. Антикоммутатор $\frac{1}{2}\{\Psi{}^{+},\Phi\}$
удовлетворяет всем требованиям к скалярному произведению величин $\Psi$
и $\Phi$, и его можно считать скалярным произведением $(\Psi,\Phi)$.
Поэтому (1) и (2) можно переписать как

\[
(\theta^{b}(p),\theta^{a}(p'))=(\frac{\partial}{\partial\theta^{a}(p')},\frac{\partial}{\partial\theta^{b}(p)})=\frac{1}{2}\delta_{b}^{a}\delta(p-p'),
\]
\[
(\theta^{a}(p),\frac{\partial}{\partial\theta^{b}(p')})=(\frac{\partial}{\partial\theta^{a}(p)},\theta^{b}(p'))=0.
\]

Множитель $\frac{1}{2}$ введен для того, чтобы далее получилось $(1,1)=1$.

Рассмотрим двусторонний модуль над алгеброй с единицей, в которой
величины $\theta^{a}(p)$ и $\partial/\partial\theta^{a}(p)$ являются
образующими. Будем предполагать, что в состав модуля входит 1, и модуль
предлагается рассматривать как бимодуль над CAR-алгеброй, с возможностью
умножения образующих алгебры на элементы модуля как слева, так и справа.
Базис модуля состоит из 1 и величин 
\begin{equation}
\Phi_{\lambda_{1}...\lambda_{\text{a}}}^{\chi_{1}...\chi_{b}}=\frac{\partial}{\partial\theta^{\lambda_{1}}(p)}...\frac{\partial}{\partial\theta^{\lambda_{a}}(p')}\theta^{\chi_{1}}(p'')...\theta^{\chi_{b}}(p''').
\end{equation}

В силу (1), (2) мы можем считать, что в (3) величины $\partial/\partial\theta^{\lambda}(p)$
всегда стоят слева от $\theta^{\chi}(p)$. Действие элемента $\theta^{\chi}(p)$
из алгебры на единицу модуля как слева, так и справа дает элемент
$\theta^{\chi}(p)$ из модуля: $\theta^{\chi}(p)1=1\theta^{\chi}(p)=\theta^{\chi}(p)$.
В результате действия оператора $\theta^{\chi}(p)$ из алгебры слева
или справа на любой элемент $\Phi$ модуля, не содержащий $\theta^{\chi}(p)$
или $\partial/\partial\theta^{\chi}(p)$, результат соответствует
дополнительному элементу $\theta^{\chi}(p)$ в качестве элемента каждого
монома в $\Phi$ с учетом изменения знака перед мономом при нечетной
перестановке этого элемента слева или справа на соответствующее место.
А при наличии в $\Phi$ $\text{величин }\theta^{\chi}(p)$ или $\partial/\partial\theta^{\chi}(p)$
обмен местами или обнуление производится с учетом (1) и (2). Поэтому
элементы $\theta^{\chi}(p)$ из алгебры и из модуля можно обозначать
одним символом. В то же время возможны два варианта действия на единицу
модуля элементов $\partial/\partial\theta^{\lambda}(p)$: в первом
не только $\theta^{\chi}(p)$, но и $\partial/\partial\theta^{\lambda}(p)$
являются образующими модуля, и $\frac{\partial}{\partial\theta^{\lambda}(p)}1=1\frac{\partial}{\partial\theta^{\lambda}(p)}=\frac{\partial}{\partial\theta^{\lambda}(p)}$
, во втором варианте $\frac{\partial}{\partial\theta^{\lambda}(p)}1=1\frac{\partial}{\partial\theta^{\lambda}(p)}=0$.
Мы будем рассматривать первый вариант, так как в нем удалось построить
как нетривиальное вакуумное состояние (отличное от 1), так и супералгебраическую
форму преобразований Лоренца {[}16{]}.

Обобщение скалярного произведения для произвольных элементов модуля
вида (3), составленных из грассмановых плотностей и производных по
ним, осуществляется по правилам, аналогичным действию полиформ на
поливекторы в дифференциальной геометрии, где аналогом касательных
векторов служат грассмановы плотности $\partial/\partial\theta^{a}(p)$,
а дифференциальных 1-форм \textendash{} грассмановы плотности $\theta^{a}(p)$.
Единственное заметное отличие заключается в появлении в аналогах полиформ
помимо множителей вида $\theta^{a}(p)$ множителей вида $\partial/\partial\theta^{b}(p')$,
и появлении в аналогах поливекторов помимо $\partial/\partial\theta^{a}(p)$
множителей вида $\theta^{b}(p')$. Требования $\theta^{a}(p)^{+}=\partial/\partial\theta^{a}(p)$
и $(\partial/\partial\theta^{a}(p))^{+}=\theta^{a}(p)$ приводят к
соотношениям

\[
(\theta^{b}(p),\theta^{a}(p'))=(1,\frac{\partial}{\partial\theta^{b}(p)}\theta^{a}(p'))=\frac{1}{2}\delta_{b}^{a}\delta(p-p'),
\]

\[
(\frac{\partial}{\partial\theta^{a}(p')},\frac{\partial}{\partial\theta^{b}(p)})=(1,\theta^{a}(p')\frac{\partial}{\partial\theta^{b}(p)})=\frac{1}{2}\delta_{b}^{a}\delta(p-p'),
\]

\[
(1,\frac{\partial}{\partial\theta^{b}(p)}\theta^{a}(p')-\theta^{a}(p')\frac{\partial}{\partial\theta^{b}(p)})=\frac{1}{2}\delta_{b}^{a}\delta(p-p')-\frac{1}{2}\delta_{b}^{a}\delta(p-p')=0,
\]

\[
(1,\frac{\partial}{\partial\theta^{b}(p)}\theta^{a}(p')+\theta^{a}(p')\frac{\partial}{\partial\theta^{b}(p)})=(1,1)(\frac{1}{2}\delta_{b}^{a}\delta(p-p')+\frac{1}{2}\delta_{b}^{a}\delta(p-p'))=\delta_{b}^{a}\delta(p-p'),
\]

\[
(1,\frac{\partial}{\partial\theta^{b}(p)}\theta^{a}(p')+\theta^{a}(p')\frac{\partial}{\partial\theta^{b}(p)})=(1,1)(\frac{1}{2}\delta_{b}^{a}\delta(p-p')+\frac{1}{2}\delta_{b}^{a}\delta(p-p'))=\delta_{b}^{a}\delta(p-p'),
\]

То есть $(1,1)=1$, а скалярное произведение 1 и $\frac{\partial}{\partial\theta^{b}(p)}\theta^{a}(p')-\theta^{a}(p')\frac{\partial}{\partial\theta^{b}(p)}$
равно нулю. 

Рассмотрим скалярное произведение $(\Phi,\Psi)$ величин $\Phi=\Phi_{\lambda_{1}...\lambda_{\text{a}}}^{\chi_{1}...\chi_{b}}$и
$\Psi=\Psi_{\mu_{1}...\mu_{\text{c}}}^{\eta_{1}...\eta_{d}}$вида
(3). 

\[
(\Phi,\Psi)=(\frac{\partial}{\partial\theta^{\lambda_{1}}(p)}...\frac{\partial}{\partial\theta^{\lambda_{a}}(p')}\theta^{\chi_{1}}(p'')...\theta^{\chi_{b}}(p'''),\Psi).
\]

Перенося из левой части скалярного произведения в правую множители
$\partial/\partial\theta^{a}(p)$, получаем величины $\theta^{a}(p)$,
перемноженные в обратном порядке, а перенося множители $\theta^{b}(p)$,
получаем величины $\partial/\partial\theta^{b}(p)$, перемноженные
в обратном порядке. Результат переноса обозначим как $\Phi^{+}$:

\[
(\Phi,\Psi)=(1,\frac{\partial}{\partial\theta^{\chi_{b}}(p''')}...\frac{\partial}{\partial\theta^{\chi_{1}}(p'')}\theta^{\lambda_{a}}(p')...\theta^{\lambda_{1}}(p)\Psi)=(1,\Phi^{+}\Psi).
\]

Введем общее правило нахождения такого скалярного произведения аналогично
правилу действия полиформ на поливекторы с учетом упомянутых ранее
особенностей. В нем величины $\partial/\partial\theta^{a}(p)$ и $\theta^{b}(p)$,
входящие в состав $\Phi^{+}\Psi$, действуют по модифицированному
правилу Лейбница на стоящие справа от них элементы \textendash{} сворачиваются
с ними. Если между сворачиваемыми элементами стоит нечетное число
множителей, перед соответствующим слагаемым указывается знак минус,
если четное (в том числе ноль), указывается знак плюс. Свертки элементов
$\partial/\partial\theta^{a}(p)$ с $\partial/\partial\theta^{b}(p')$
и $\theta^{a}(p)$ с $\theta^{b}(p')$ равны нулю. Свертка $\partial/\partial\theta^{a}(p)$
с $\theta^{b}(p')$ дает $\frac{1}{2}\delta_{a}^{b}\delta(p-p')$,
такой же результат дает свертка $\theta^{b}(p')$ с $\partial/\partial\theta^{a}(p)$.
Скалярное произведение $(1,1)=1,$ а скалярное произведение 1 с любыми
другими оставшимися после сверток элементами дает 0. 

Если $\Phi^{+}$ является аналогом полиформы, а $\Psi$ аналогом поливектора,
данное правило идентично правилу действия полиформы на поливектор.

Легко проверить, что введенное правило согласуется со всеми приведенными
ранее формулами и позволяет находить скалярные произведения любых
элементов вида (3). Например, в соответствии с этим правилом получаем,
что

\[
(\frac{\partial}{\partial\theta^{b}(p_{1})}\theta^{a}(p_{2}),\frac{\partial}{\partial\theta^{d}(p_{3})}\theta^{c}(p_{4}))=\frac{1}{4}\delta_{b}^{a}\delta_{d}^{c}\delta(p_{1}-p_{2})\delta(p_{3}-p_{4})+\frac{1}{4}\delta_{d}^{b}\delta_{a}^{c}\delta(p_{1}-p_{3})\delta(p_{2}-p_{4}).
\]

Рассмотрим теперь операторы, действующие на элементы модуля. Поскольку
модуль является бимодулем над CAR-алгеброй, и определены левое и правое
умножение образующих $\partial/\partial\theta^{\lambda}(p)$ и $\theta^{\chi}(p)$
на произвольный элемент $\Phi$ данного модуля, для любого элемента
$A$ алгебры вместо правого умножения $A$ на $\Phi$, то есть $\Phi A$,
удобно рассматривать коммутатор $[A,*]$, который далее будем обозначать
как $\hat{A}$, и который является разностью между результатами левого
и правого умножения: 
\[
\hat{A}\Phi=[A,*]\Phi=A\Phi-\Phi A.
\]

Таким образом, операторы, действующие на элементы модуля, в общем
случае будут состоять из сумм и произведений величин вида (3) и коммутаторов.
При этом 
\begin{equation}
(\hat{A}\Phi)^{+}=(A\Phi-\Phi A)^{+}=\Phi^{+}A{}^{+}-A{}^{+}\Phi^{+}=-\hat{A}^{+}\Phi^{+}.
\end{equation}

Будем искать условия, при которых у гамма-матриц Дирака $\gamma^{\mu}$
существуют супералгебраические аналоги $\hat{\gamma}^{\mu}$(реализация
клиффордовых векторов с помощью грассмановых плотностей), обладающие
такими же коммутационными соотношениями и так же себя ведущих при
эрмитовом сопряжении. Будем называть собственными супералгебраическими
преобразованиями Лоренца супералгебраическое представление псевдоортогональных
(клиффордовых) вращений псевдоевклидова пространства произвольной
размерности. Они осуществляются оператором $\hat{T}=exp(\hat{\gamma}^{\mu}\hat{\gamma}^{\nu}\varphi_{\mu\nu}/2)$,
где $\varphi_{\mu\nu}$ \textendash{} параметры поворота {[}16{]}:
\[
\Phi'=e^{\hat{\gamma}^{\mu}\hat{\gamma}^{\nu}\varphi_{\mu\nu}/2}\Phi,
\]

\[
\hat{A}'=e^{\hat{\gamma}^{\mu}\hat{\gamma}^{\nu}\varphi_{\mu\nu}/2}\hat{A}e^{-\hat{\gamma}^{\mu}\hat{\gamma}^{\nu}\varphi_{\mu\nu}/2}.
\]

Пусть оператор $\Psi(p)$ является линейной комбинацией образующих
для заданного значения $p$: 
\[
\Psi(p)=\psi_{\alpha}(p)\frac{\partial}{\partial\theta^{\alpha}(p)}+\psi_{\tau}(p)\theta^{\tau}(p).
\]

Будем рассматривать ситуацию, когда существует Лоренц-инвариантный
вектор состояния вакуума $\Phi_{V}$, то есть когда выполняется соотношение
\[
e^{\hat{\gamma}^{\mu}\hat{\gamma}^{\nu}\varphi_{\mu\nu}/2}\Phi_{V}=\Phi_{V}.
\]

При умножении оператора поля $\Psi(p)$ на $\Phi_{V}$ получаем новый
вектор состояния $\Psi(p)\Phi_{V}$, который преобразуется как 
\[
e^{\hat{\gamma}^{\mu}\hat{\gamma}^{\nu}\varphi_{\mu\nu}/2}\Psi(p)\Phi_{V}=(e^{\hat{\gamma}^{\mu}\hat{\gamma}^{\nu}\varphi_{\mu\nu}/2}\Psi(p))\Phi_{V},
\]
где скобки ограничивают область действия оператора $exp(\hat{\gamma}^{\mu}\hat{\gamma}^{\nu}\varphi_{\mu\nu}/2)$.
Поэтому в случае подобных состояний (одночастичных) можно считать,
что 
\begin{equation}
\Psi'(p)=e^{\hat{\gamma}^{\mu}\hat{\gamma}^{\nu}\varphi_{\mu\nu}/2}\Psi(p).
\end{equation}

Эрмитово сопрягая (5) и учитывая (4), получаем, что эрмитово сопряженный
оператор поля преобразуется по правилу {[}16{]} 
\begin{equation}
\Psi'(p)^{+}=e^{\hat{\gamma}^{\mu+}\hat{\gamma}^{\nu+}\varphi_{\mu\nu}/2}\Psi(p)^{+}.
\end{equation}

В случае, когда все гамма-матрицы имеют одинаковую сигнатуру, преобразование
(6) превращается в соотношение (7): 
\begin{equation}
\Psi'(p){}^{+}=e^{\hat{\gamma}^{\mu}\hat{\gamma}^{\nu}\varphi_{\mu\nu}/2}\Psi(p){}^{+}.
\end{equation}

Которое означает, что сопряженные операторы поля при преобразованиях
Лоренца преобразуются ковариантно. Если же среди гамма-матриц имеются
матрицы с разной сигнатурой и, например, $\gamma^{0+}=\gamma^{0}$,
а $\gamma^{k+}=-\gamma^{k},\,k=1,2,3$, то эрмитово сопряженные операторы
поля при преобразованиях Лоренца преобразуются нековариантно, так
как знак в показателе экспоненты меняется на противоположный: 
\begin{equation}
\Psi'(p){}^{+}=e^{-\hat{\gamma}^{0}\hat{\gamma}^{k}\varphi_{0k}/2}\Psi(p){}^{+}.
\end{equation}

Требование замены преобразований (8) сопряженных спиноров на Лоренц-ковариантные
приводит к необходимости введения обобщенного дираковского сопряжения.

\section{Обобщенное дираковское сопряжение операторов для сигнатуры пространства-времени
(1, n-1)}

Назовем обобщенным дираковским сопряжением оператора $\Psi$ величину
\begin{equation}
\bar{\Psi}=(M\Psi)^{+},
\end{equation}
где $M$ \textendash{} некая матрица (оператор) в супералгебраическом
представлении, а $\Psi$ является линейной комбинацией величин $\Phi_{\lambda_{1}...\lambda_{\text{a}}}^{\chi_{1}...\chi_{b}}$
вида (3). Если оператор $M$ единичный, обобщенное дираковское сопряжение
является эрмитовым сопряжением, а если $M=\hat{\gamma}^{0}$, обобщенное
дираковское сопряжение является супералгебраическим аналогом обычного
дираковского сопряжения. В дальнейшем для краткости слово ``обобщенное\textquotedblright{}
будем опускать. Рассмотрим возможные варианты сопряжений (9), обеспечивающие
Лоренц-ковариантность преобразований величины $\bar{\Psi}$, то есть
такие, при которых выполняется условие (10): 
\begin{equation}
\bar{\Psi}'(p)=e^{\hat{\gamma}^{0}\hat{\gamma}^{k}\varphi_{0k}/2}\bar{\Psi}(p).
\end{equation}

Пусть имеется n-мерное псевдоевклидово пространство с сигнатурой $(p,q)$,
где $p+q=n$. Обозначение $p$, относящееся к сигнатуре, по контексту
легко отличить от обозначения $p$, относящегося к параметру грассмановой
плотности (энергии-импульсу), поэтому мы будем использовать для них
одну и ту же букву.

Обозначим базисные клиффордовы векторы с положительной сигнатурой
как $\hat{\gamma}_{+}^{1},\hat{\gamma}_{+}^{2},\ldots,\hat{\gamma}_{+}^{p}$,
а с отрицательной сигнатурой как $\hat{\gamma}_{-}^{1},\hat{\gamma}_{-}^{2},\ldots,\hat{\gamma}_{-}^{q}$.
Далее будем исходить из того, что оператор $M$ в (9) является элементом
алгебры Клиффорда, и поэтому построен из сумм и произведений базисных
клиффордовых векторов с некими комплексными или вещественными коэффициентами.
В соответствии с периодичностью Картана-Ботта возможны три варианта
матричных представлений алгебр Клиффорда \textendash{} вещественные,
комплексные и кватернионные {[}1{]}, {[}10{]}. В случае если матричная
алгебра вещественная, элементы матриц вещественные, в случае комплексной
\textendash{} комплексные. В случае кватернионной алгебры можно рассматривать
кватернионные элементы матриц, но удобнее рассматривать комплексные
матрицы с удвоенным количеством строк и столбцов. Супералгебраическое
представление матриц полностью повторяет эти варианты.

В более общем случае оператор $M$ может являться оператором обобщенной
матричной алгебры {[}14{]}, {[}15{]} и приводить к преобразованиям
суперсимметрии {[}15{]}, однако такой вариант требует отдельного рассмотрения.

Рассмотрим случай, когда один базисный клиффордов вектор $\hat{\gamma}_{+}^{1}$
имеет положительную сигнатуру, а остальные $\hat{\gamma}_{-}^{1},\hat{\gamma}_{-}^{2},\ldots,\hat{\gamma}_{-}^{q}$
\textendash{} отрицательную, где $q=n-1$. Представим $M$ в виде

\begin{flushleft}
\[
\begin{array}{ccc}
M & = & M_{0}(\hat{\gamma}_{-}^{2},\ldots,\hat{\gamma}_{-}^{q})+\hat{\gamma}_{+}^{1}M_{+}^{1}(\hat{\gamma}_{-}^{2},\ldots,\hat{\gamma}_{-}^{q})\,+\,\,\,\,\,\,\,\,\,\,\,\,\,\,\,\,\,\\
 &  & \hat{\gamma}_{-}^{1}M_{-}^{1}(\hat{\gamma}_{-}^{2},\ldots,\hat{\gamma}_{-}^{q})+\hat{\gamma}_{+}^{1}\hat{\gamma}_{-}^{1}M_{+-}^{11}(\hat{\gamma}_{-}^{2},\ldots,\hat{\gamma}_{-}^{q}),
\end{array}
\]
где в скобках указаны базисные клиффордовы векторы, от которых может
зависеть соответствующая величина $M_{\cdots}^{\cdots}$, то есть
в состав которой могут входить слагаемые с мономами, содержащими данные
векторы. Будем считать параметры в скобках частью обозначения величин.
Поэтому, например, $M_{0}(\hat{\gamma}_{-}^{2},\ldots,\hat{\gamma}_{-}^{q})$
и $M_{0}(\hat{\gamma}_{-}^{3},\ldots,\hat{\gamma}_{-}^{q})$ это разные
величины $M_{0}$.
\par\end{flushleft}

Тогда из (9) следует 
\begin{equation}
\begin{array}{ccc}
\bar{\Psi} & = & (M_{0}(\hat{\gamma}_{-}^{2},\ldots,\hat{\gamma}_{-}^{q})^{+}-\hat{\gamma}_{+}^{1}M_{+}^{1}(\hat{\gamma}_{-}^{2},\ldots,\hat{\gamma}_{-}^{q})^{+}\,\,+\,\,\,\,\,\,\,\,\,\,\,\,\,\,\,\,\,\,\,\,\,\,\\
 &  & \hat{\gamma}_{-}^{1}M_{-}^{1}(\hat{\gamma}_{-}^{2},\ldots,\hat{\gamma}_{-}^{q})^{+}-\hat{\gamma}_{+}^{1}\hat{\gamma}_{-}^{1}M_{+-}^{11}(\hat{\gamma}_{-}^{2},\ldots,\hat{\gamma}_{-}^{q})^{+})\Psi^{+}.
\end{array}
\end{equation}

Рассмотрим преобразование 
\[
\Psi'(p)=e^{\hat{\gamma}_{+}^{1}\hat{\gamma}_{-}^{1}\varphi/2}\Psi(p),
\]
\[
\Psi'(p)^{+}=e^{-\hat{\gamma}_{+}^{1}\hat{\gamma}_{-}^{1}\varphi/2}\Psi(p)^{+},
\]
где $\varphi$ \textendash{} произвольный вещественный параметр (угол
поворота в плоскости $\hat{\gamma}_{+}^{1},\hat{\gamma}_{-}^{1}$).

При таком преобразовании первое и последнее слагаемые, получаемые
в (11) после раскрытия скобок, если они ненулевые, преобразуются нековариантно.
Поэтому $M_{0}(\hat{\gamma}_{-}^{2},\ldots,\hat{\gamma}_{-}^{q})=M_{+-}^{11}(\hat{\gamma}_{-}^{2},\ldots,\hat{\gamma}_{-}^{q})=0$,
и получаем 
\[
M=\hat{\gamma}_{+}^{1}M_{+}^{1}(\hat{\gamma}_{-}^{2},\ldots,\hat{\gamma}_{-}^{q})+\hat{\gamma}_{-}^{1}M_{-}^{1}(\hat{\gamma}_{-}^{2},\ldots,\hat{\gamma}_{-}^{q}),
\]
\begin{equation}
\bar{\Psi}=(-\hat{\gamma}_{+}^{1}M_{+}^{1}(\hat{\gamma}_{-}^{2},\ldots,\hat{\gamma}_{-}^{q})^{+}+\hat{\gamma}_{-}^{1}M_{-}^{1}(\hat{\gamma}_{-}^{2},\ldots,\hat{\gamma}_{-}^{q})^{+})\Psi^{+}.
\end{equation}
Введем обозначение 
\[
\hat{\gamma}=\hat{\gamma}_{+}^{1}\hat{\gamma}_{-}^{1}\ldots\hat{\gamma}_{-}^{q}.
\]

Если $n=2,\,p=1,\,q=1$, получаем $\hat{\gamma}=\hat{\gamma}_{+}^{1}\hat{\gamma}_{-}^{1}$,
$\hat{\gamma}^{2}=1$, $\hat{\gamma}_{-}^{1}=\hat{\gamma}_{+}^{1}\hat{\gamma}$.
Поэтому 
\begin{equation}
M=\hat{\gamma}_{+}^{1}(M_{+}^{1}+\hat{\gamma}M_{-}^{1}).
\end{equation}

Если $n>2,\,p=1,\,q>1$, разложим в (12) величины $M_{+}^{1}(\hat{\gamma}_{-}^{2},\ldots,\hat{\gamma}_{-}^{q})$
и $M_{-}^{1}(\hat{\gamma}_{-}^{2},\ldots,\hat{\gamma}_{-}^{q})$:
\[
M_{+}^{1}(\hat{\gamma}_{-}^{2},\ldots,\hat{\gamma}_{-}^{q})=M_{+}^{1}(\hat{\gamma}_{-}^{3},\ldots,\hat{\gamma}_{-}^{q})+\hat{\gamma}_{-}^{2}M_{+-}^{12}(\hat{\gamma}_{-}^{3},\ldots,\hat{\gamma}_{-}^{q}),
\]
\[
M_{-}^{1}(\hat{\gamma}_{-}^{2},\ldots,\hat{\gamma}_{-}^{q})=M_{-}^{1}(\hat{\gamma}_{-}^{3},\ldots,\hat{\gamma}_{-}^{q})+\hat{\gamma}_{-}^{2}M_{--}^{12}(\hat{\gamma}_{-}^{3},\ldots,\hat{\gamma}_{-}^{q}).
\]
\begin{equation}
\begin{array}{ccc}
\bar{\Psi} & = & (-\hat{\gamma}_{+}^{1}M_{+}^{1}(\hat{\gamma}_{-}^{3},\ldots,\hat{\gamma}_{-}^{q})^{+}-\hat{\gamma}_{+}^{1}\hat{\gamma}_{-}^{2}M_{+-}^{12}(\hat{\gamma}_{-}^{3},\ldots,\hat{\gamma}_{-}^{q})^{+}+\\
 &  & (\hat{\gamma}_{-}^{1}M_{-}^{1}(\hat{\gamma}_{-}^{3},\ldots,\hat{\gamma}_{-}^{q})^{+}+\hat{\gamma}_{-}^{1}\hat{\gamma}_{-}^{2}M_{--}^{12}(\hat{\gamma}_{-}^{3},\ldots,\hat{\gamma}_{-}^{q})^{+})\Psi^{+}.
\end{array}
\end{equation}

Рассмотрим преобразование 
\begin{equation}
\Psi'(p)=e^{\hat{\gamma}_{+}^{1}\hat{\gamma}_{-}^{2}\varphi/2}\Psi(p),
\end{equation}
\[
\Psi'(p)^{+}=e^{-\hat{\gamma}_{+}^{1}\hat{\gamma}_{-}^{2}\varphi/2}\Psi(p)^{+}.
\]

При нем второе и третье слагаемые в (14) после раскрытия скобок, если
они ненулевые, преобразуются нековариантно. Поэтому $M_{+-}^{12}(\hat{\gamma}_{-}^{3},\ldots,\hat{\gamma}_{-}^{q})=M_{-}^{1}(\hat{\gamma}_{-}^{3},\ldots,\hat{\gamma}_{-}^{q})=0$,
и 
\[
M=\hat{\gamma}_{+}^{1}M_{+}^{1}(\hat{\gamma}_{-}^{3},\ldots,\hat{\gamma}_{-}^{q})+\hat{\gamma}_{-}^{1}\hat{\gamma}_{-}^{2}M_{--}^{12}(\hat{\gamma}_{-}^{3},\ldots,\hat{\gamma}_{-}^{q}),
\]
\begin{equation}
\bar{\Psi}=(-\hat{\gamma}_{+}^{1}M_{+}^{1}(\hat{\gamma}_{-}^{3},\ldots,\hat{\gamma}_{-}^{q})^{+}+\hat{\gamma}_{-}^{1}\hat{\gamma}_{-}^{2}M_{--}^{12}(\hat{\gamma}_{-}^{3},\ldots,\hat{\gamma}_{-}^{q})^{+})\Psi^{+}.
\end{equation}

При рассмотрении поворотов в пространстве с образующими $\hat{\gamma}_{+}^{1},\,\hat{\gamma}_{-}^{1},\,\hat{\gamma}_{-}^{2}$
помимо преобразования (15) возможно преобразование 
\[
\Psi'(p)=e^{\hat{\gamma}_{-}^{1}\hat{\gamma}_{-}^{2}\varphi/2}\Psi(p),
\]
\[
\Psi'(p)^{+}=e^{\hat{\gamma}_{-}^{1}\hat{\gamma}_{-}^{2}\varphi/2}\Psi(p)^{+}.
\]

Очевидно, что оно ковариантно преобразует величину $\bar{\Psi}$,
задаваемую выражением (16). Продолжая процесс для $\hat{\gamma}_{-}^{3},\ldots,\hat{\gamma}_{-}^{q}$
, получаем в итоге 
\begin{equation}
M=\hat{\gamma}_{+}^{1}M_{+}^{1}+\hat{\gamma}_{-}^{1}\hat{\gamma}_{-}^{2}\ldots\hat{\gamma}_{-}^{q}M_{--\ldots-}^{12\ldots q}.
\end{equation}

Поскольку в этом случае $\hat{\gamma}=\hat{\gamma}_{+}^{1}\hat{\gamma}_{-}^{1}\hat{\gamma}_{-}^{2}\ldots\hat{\gamma}_{-}^{q}$,
и $\hat{\gamma}_{-}^{1}\hat{\gamma}_{-}^{2}\ldots\hat{\gamma}_{-}^{q}=\hat{\gamma}_{+}^{1}\hat{\gamma}$,
можно переписать (17) в виде 
\begin{equation}
M=\hat{\gamma}_{+}^{1}(M_{+}^{1}+\hat{\gamma}M_{--\ldots-}^{12\ldots q}).
\end{equation}

Из (13) видно, что формула (18) применима и для $n=2,\,p=1,\,q=1$.
Поэтому она верна для $n\geq2,\,p=1,\,q\geq1$. Свойства клиффордова
числа $\hat{\gamma}=\hat{\gamma}_{+}^{1}\hat{\gamma}_{-}^{1}\hat{\gamma}_{-}^{2}\ldots\hat{\gamma}_{-}^{q}$
заметно отличаются для четномерных $n=2m$ и нечетномерных $n=2m+1$
пространств и при разных размерностях. В случае нечетномерных пространств
$\hat{\gamma}$ коммутирует со всеми базисными клиффордовыми векторами
и потому является, наряду с единицей, образующей центра соответствующей
клиффордовой алгебры. Для рассматриваемого случая $(p,q)=(1,n-1)$
при $n=3$ имеем $(\hat{\gamma})^{2}=-1$, то есть $\hat{\gamma}$
играет роль клиффордовой мнимой единицы. При $n=5$ получается $(\hat{\gamma})^{2}=1$.

В общем случае для нечетномерных пространств размерности $n=2m+1$
\[
(\hat{\gamma})^{2}=(-1)^{m}.
\]

В случае четномерных пространств $n=2m$ величина $\hat{\gamma}$
антикоммутирует со всеми базисными клиффордовыми векторами и потому
с точностью до сигнатуры является аналогом известной матрицы Дирака
$\gamma^{5}$ . При этом 
\[
(\hat{\gamma})^{2}=(-1)^{m-1}.
\]

Для случая $n=4$ по определению $\hat{\gamma}^{5}=i\hat{\gamma}_{+}^{1}\hat{\gamma}_{-}^{1}\hat{\gamma}_{-}^{2}\hat{\gamma}_{-}^{3}=i\hat{\gamma}$.

\section{Обобщенное дираковское сопряжение для произвольной сигнатуры пространства-времени}

В случае сигнатуры $(p,q)=(n,0)$ тем же методом получаем 
\[
M=\hat{\gamma}_{+}^{1}\hat{\gamma}_{+}^{2}\ldots\hat{\gamma}_{+}^{n}M_{+\ldots+}^{12\ldots n}+M_{-},
\]
а в случае сигнатуры $(p,q)=(0,n)$ получаем 
\[
M=M_{+}+\hat{\gamma}_{-}^{1}\hat{\gamma}_{-}^{2}\ldots\hat{\gamma}_{-}^{n}M_{--\ldots-}^{12\ldots n}
\]

Рассмотрение случая $p=2,q=1$ совершенно аналогично случаю $p=1,\,q=2$,
в результате получается 
\[
M=\hat{\gamma}_{+}^{1}\hat{\gamma}_{+}^{2}M_{++}^{12}+\hat{\gamma}_{-}^{1}M_{-}^{1}.
\]

В случае $p=2,q=n-2>1$ справедливы все рассуждения, приводящие к
формуле (17), только величины $M_{+}^{1}$ и $M_{--\ldots-}^{12\ldots q}$
становятся зависимыми от $\hat{\gamma}_{+}^{2}$. Поэтому выполняется
соотношение 
\[
M=\hat{\gamma}_{+}^{1}M_{+}^{1}(\hat{\gamma}_{+}^{2})+\hat{\gamma}_{-}^{1}\hat{\gamma}_{-}^{2}\ldots\hat{\gamma}_{-}^{q}M_{--\ldots-}^{12\ldots q}(\hat{\gamma}_{+}^{2}).
\]

Совершенно аналогично предыдущим выкладкам получаем 
\[
M=\hat{\gamma}_{+}^{1}M_{+}^{1}+\hat{\gamma}_{+}^{1}\hat{\gamma}_{+}^{2}M_{++}^{12}+\hat{\gamma}_{-}^{1}\hat{\gamma}_{-}^{2}\ldots\hat{\gamma}_{-}^{q}M_{--\ldots-}^{12\ldots q}+\hat{\gamma}_{+}^{2}\hat{\gamma}_{-}^{1}\hat{\gamma}_{-}^{2}\ldots\hat{\gamma}_{-}^{q}M_{+--\ldots-}^{212\ldots q},
\]
\begin{equation}
\begin{array}{ccc}
\bar{\Psi} & = & -\hat{\gamma}_{+}^{1}(M_{+}^{1})^{+}\Psi^{+}+\hat{\gamma}_{+}^{1}\hat{\gamma}_{+}^{2}(M_{++}^{12})^{+}\Psi^{+}+\hat{\gamma}_{-}^{1}\hat{\gamma}_{-}^{2}\ldots\hat{\gamma}_{-}^{q}(M_{--\ldots-}^{12\ldots q})^{+}\Psi^{+}-\\
 &  & \hat{\gamma}_{+}^{2}\hat{\gamma}_{-}^{1}\hat{\gamma}_{-}^{2}\ldots\hat{\gamma}_{-}^{q}(M_{+--\ldots-}^{212\ldots q})^{+}\Psi^{+}.
\end{array}
\end{equation}

Рассмотрим преобразование 
\[
\Psi'(p)=e^{\hat{\gamma}_{+}^{1}\hat{\gamma}_{+}^{2}\varphi/2}\Psi(p),
\]
\[
\Psi'(p)^{+}=e^{\hat{\gamma}_{+}^{1}\hat{\gamma}_{+}^{2}\varphi/2}\Psi(p)^{+}.
\]

При нем первое и последнее слагаемое в (19), если они ненулевые, преобразуются
нековариантно. Поэтому $M_{+}^{1}=M_{+--\ldots-}^{212\ldots q}=0$,
и 
\[
M=\hat{\gamma}_{+}^{1}\hat{\gamma}_{+}^{2}M_{++}^{12}+\hat{\gamma}_{-}^{1}\hat{\gamma}_{-}^{2}\ldots\hat{\gamma}_{-}^{q}M_{--\ldots-}^{12\ldots q}.
\]

Обозначим по аналогии со случаем с одной времениподобной осью 
\[
\hat{\gamma}=\hat{\gamma}_{+}^{1}\hat{\gamma}_{+}^{2}\hat{\gamma}_{-}^{1}\ldots\hat{\gamma}_{-}^{q}.
\]

Тогда $\hat{\gamma}_{-}^{1}\ldots\hat{\gamma}_{-}^{q}=-\hat{\gamma}_{+}^{1}\hat{\gamma}_{+}^{2}\hat{\gamma}$,
поэтому 
\[
M=\hat{\gamma}_{+}^{1}\hat{\gamma}_{+}^{2}(M_{++}^{12}-\hat{\gamma}M_{--\ldots-}^{12\ldots q}).
\]

С учетом того, что $\hat{\gamma}^{+}=\hat{\gamma}$, получаем 
\begin{equation}
\bar{\Psi}=\hat{\gamma}_{+}^{1}\hat{\gamma}_{+}^{2}((M_{++}^{12})^{+}-\hat{\gamma}(M_{--\ldots-}^{12\ldots q})^{+})\Psi^{+}.
\end{equation}

При преобразованиях 
\[
\Psi'(p)=e^{\hat{\gamma}_{+}^{i}\hat{\gamma}_{-}^{j}\varphi/2}\Psi(p),
\]
\[
\Psi'(p)^{+}=e^{-\hat{\gamma}_{+}^{i}\hat{\gamma}_{-}^{j}\varphi/2}\Psi(p)^{+}.
\]
величина $\bar{\Psi}$, задаваемая формулой (20), преобразуется ковариантно.
Поэтому при всех возможных преобразованиях Лоренца $\bar{\Psi}$ преобразуется
ковариантно.

Совершенно аналогичным образом получаем формулы для $p=3,\,q=4$,
и так далее. В итоге получаем формулу (21) 
\begin{equation}
M=\hat{\gamma}_{+}^{1}\ldots\hat{\gamma}_{+}^{p}M_{+}+\hat{\gamma}_{-}^{1}\ldots\hat{\gamma}_{-}^{q}M_{-},
\end{equation}
справедливую для всех значений $p$ и $q$. Если в (21) $p=0$, то
$M=M_{+}+\hat{\gamma}_{-}^{1}\ldots\hat{\gamma}_{-}^{q}M_{-}$, а
если $q=0$, то $M=\hat{\gamma}_{+}^{1}\ldots\hat{\gamma}_{+}^{p}M_{+}+M_{-}$.

Будем помечать величины $M$ для сигнатуры $(p,q)$ как $M$$_{p,q}$.
При этом (21) приобретет окончательный вид 
\begin{equation}
M_{p,q}=\hat{\gamma}_{+}^{1}\ldots\hat{\gamma}_{+}^{p}M_{+,p,q}+\hat{\gamma}_{-}^{1}\ldots\hat{\gamma}_{-}^{q}M_{-,p,q}.
\end{equation}

В случае нечетного числа измерений $n=p+q$ величина 
\[
\hat{\gamma}=\hat{\gamma}_{+}^{1}\ldots\hat{\gamma}_{+}^{p}\hat{\gamma}_{-}^{1}\ldots\hat{\gamma}_{-}^{q}.
\]
принадлежит центру алгебры. В результате (22) можно переписать в виде
\begin{equation}
M_{p,q}=\hat{\gamma}_{+}^{1}\ldots\hat{\gamma}_{+}^{p}M_{+}=\hat{\gamma}_{-}^{1}\ldots\hat{\gamma}_{-}^{q}M_{-}.
\end{equation}
где величины $M_{+}$ и $M_{-}$ являются коммутирующими со всеми
элементами алгебры гиперкомплексными константами, в состав которых
может входить величина $\hat{\gamma}$.

Необходимо отметить еще один момент: в приведенных доказательствах
рассматривалась ковариантность относительно преобразований Лоренца.
Если же, как будет показано далее, имеются дополнительные по отношению
к теории Дирака базисные клиффордовы векторы, соответствующие внутренним
степеням свободы спинора, величины $M_{+,p,q}$ и $M_{-,p,q}$ в (22),
а также $M_{+}$ и $M_{-}$ в (23) могут содержать слагаемые с различными
произведениями этих дополнительных векторов.

\section{Преобразования, порожденные четырьмя грассмановыми плотностями и
производными по ним}

Пусть у нас имеется четыре грассмановы переменные (плотности) $\theta^{\alpha}(p)$
и соответствующие им производные $\partial/\partial\theta^{\alpha}(p)$,
которые также являются грассмановыми переменными. Вместе они образуют
CAR-алгебру (1), (2).

Мы уже знаем, что с помощью четырех таких переменных и производных
по ним можно образовать базисные клиффордовы векторы $\hat{\gamma}^{0}$,
$\hat{\gamma}^{1}$,$\hat{\gamma}^{2}$,$\hat{\gamma}^{3}$, $\hat{\gamma}^{5}$,
являющиеся аналогами соответствующих гамма-матриц Дирака {[}16{]}.
Встает вопрос, исчерпывается ли ими набор базисных клиффордовых векторов,
которые можно образовать из этих переменных. Для ответа на него рассмотрим
все возможные линейные преобразования этих переменных в области около
$p=0$ при действии на них операторов, сохраняющих CAR-алгебру (1),
(2). Поскольку с помощью эрмитова сопряжения грассмановых плотностей
получаются производные по ним и наоборот, такой набор будет преобразовываться
нековариантно. То есть не будут существовать операторы, действующие
как на грассмановы плотности, так и на производные по ним. Эта проблема
может быть решена использованием дираковского сопряжения вместо эрмитова,
так как в супералгебраическом представлении дираковски сопряженные
величины, как мы ранее видели, при псевдоортогональных вращениях преобразуются
ковариантно.

Величины после преобразования будем помечать штрихами. В данном разделе
будем использовать $M=\hat{\gamma}^{0}$, то есть супералгебраический
аналог обычного дираковского сопряжения $\bar{\Psi}=(\hat{\gamma}^{0}\Psi)^{+}$.

Поскольку $\hat{\gamma}^{0}\partial/\partial\theta^{i}(p)=\partial/\partial\theta^{i}(p)$,
для удобства будем рассматривать преобразования величин $\partial/\partial\theta^{i}(p)$
как первичные, а преобразования величин $\theta^{i}(p)$ как дираковски
сопряженных к ним, то есть будем считать $\theta^{i}(p)'=\overline{\partial/\partial\theta^{i}(p)}'=(\hat{\gamma}^{0}\partial'/\partial\theta^{i}(p))^{+}$.
Тогда получаем формулу для преобразованных производных 
\begin{equation}
b_{\alpha}(p')=\frac{\partial'}{\partial\theta^{\alpha}(p')}=\frac{\partial}{\partial\theta^{\alpha}(p')}+\delta k_{\beta}^{\alpha}\frac{\partial}{\partial\theta^{\beta}(p')}+\delta k_{\beta+4}^{\alpha}\theta^{\beta}(p'),
\end{equation}
а для преобразованных грассмановых плотностей 
\begin{equation}
\overline{b_{\alpha}(p')}=\theta^{\alpha}(p')'=\theta^{\alpha}(p')+(\delta k_{\beta}^{\alpha})^{*}\theta^{\beta}(p')-(\delta k_{\beta+4}^{\alpha})^{*}\frac{\partial}{\partial\theta^{\beta}(p')},
\end{equation}
где $\alpha,\beta=1,2,3,4$, величины $\delta k_{\beta}^{\alpha}$
\textendash{} бесконечно малые комплексные коэффициенты, а символ
{*} означает комплексное сопряжение.

Сохранение соотношений CAR-алгебры приводит к ряду соотношений между
коэффициентами. Из того, что 
\begin{equation}
\{\frac{\partial'}{\partial\theta^{\alpha}(p)},\frac{\partial'}{\partial\theta^{\beta}(p')}\}=0,
\end{equation}
при $\alpha=\beta$ следует 
\begin{equation}
\delta k_{<\alpha>+4}^{<\alpha>}\{\frac{\partial}{\partial\theta^{<\alpha>}(p)},\theta^{<\alpha>}(p')\}=\delta k_{<\alpha>+4}^{<\alpha>}\delta(p-p')=0.
\end{equation}

В (27) и далее по индексам в треугольных скобках нет суммирования.
Из (27) следует $\delta k_{5}^{1}=\delta k_{6}^{2}=\delta k_{7}^{3}=\delta k_{8}^{4}=0$.

Аналогично, из (26) при $\alpha=1,\,\beta=2$ следует, что 
\[
\delta k_{5}^{2}\{\frac{\partial}{\partial\theta^{1}(p)},\theta^{1}(p')\}+\delta k_{6}^{1}\{\frac{\partial}{\partial\theta^{2}(p')},\theta^{2}(p)\}=(\delta k_{5}^{2}+\delta k_{6}^{1})\delta(p-p')=0.
\]
Поэтому $\delta k_{5}^{2}+\delta k_{6}^{1}=0$, и так далее. В результате
получаем соотношения между коэффициентами: 
\begin{equation}
\begin{array}{c}
\delta k_{5}^{1}=\delta k_{6}^{2}=\delta k_{7}^{3}=\delta k_{8}^{4}=0;\\
\delta k_{1}^{1}=i\,\delta a_{1};\delta k_{2}^{2}=i\,\delta a_{2};\delta k_{3}^{3}=i\,\delta a_{3};\delta k_{4}^{4}=i\,\delta a_{4},\\
\delta k_{1}^{2}=-(\delta k_{2}^{1})^{*};\delta k_{5}^{2}=-\delta k_{6}^{1};\delta k_{1}^{3}=-(\delta k_{3}^{1})^{*};\delta k_{2}^{3}=-(\delta k_{3}^{2})^{*};\\
\delta k_{5}^{3}=-\delta k_{7}^{1};\delta k_{6}^{3}=-\delta k_{7}^{2};\delta k_{1}^{4}=-(\delta k_{4}^{1})^{*};\delta k_{2}^{4}=-(\delta k_{4}^{2})^{*};\\
\delta k_{3}^{4}=-(\delta k_{4}^{3})^{*};\delta k_{5}^{4}=-\delta k_{8}^{1};\delta k_{6}^{4}=-\delta k_{8}^{2};\delta k_{7}^{4}=-\delta k_{8}^{3},
\end{array}
\end{equation}
где $\delta a_{i}$ \textendash{} бесконечно малые вещественные параметры.

С учетом (28) соотношение (24) переходит в (29), где далее для краткости
не указывается импульс в качестве параметра: 
\begin{equation}
\begin{array}{c}
\frac{\partial'}{\partial\theta^{1}}=\frac{\partial}{\partial\theta^{1}}+i\,\delta a_{1}\frac{\partial}{\partial\theta^{1}}+\delta k_{2}^{1}\frac{\partial}{\partial\theta^{2}}+\delta k_{3}^{1}\frac{\partial}{\partial\theta^{3}}+\delta k_{4}^{1}\frac{\partial}{\partial\theta^{4}}+\delta k_{6}^{1}\theta^{2}+\delta k_{7}^{1}\theta^{3}+\delta k_{8}^{1}\theta^{4},\\
\frac{\partial'}{\partial\theta^{2}}=\frac{\partial}{\partial\theta^{2}}-(\delta k_{2}^{1})^{*}\frac{\partial}{\partial\theta^{1}}+i\,\delta a_{2}\frac{\partial}{\partial\theta^{2}}+\delta k_{3}^{2}\frac{\partial}{\partial\theta^{3}}+\delta k_{4}^{2}\frac{\partial}{\partial\theta^{4}}-\delta k_{6}^{1}\theta^{1}+\delta k_{7}^{2}\theta^{3}+\delta k_{8}^{2}\theta^{4},\\
\frac{\partial'}{\partial\theta^{3}}=\frac{\partial}{\partial\theta^{3}}-(\delta k_{3}^{1})^{*}\frac{\partial}{\partial\theta^{1}}-(\delta k_{3}^{2})^{*}\frac{\partial}{\partial\theta^{2}}+i\,\delta a_{3}\frac{\partial}{\partial\theta^{3}}+\delta k_{4}^{3}\frac{\partial}{\partial\theta^{4}}-\delta k_{7}^{1}\theta^{1}-\delta k_{7}^{2}\theta^{2}+\delta k_{8}^{3}\theta^{4},\\
\frac{\partial'}{\partial\theta^{4}}=\frac{\partial}{\partial\theta^{4}}-(\delta k_{4}^{1})^{*}\frac{\partial}{\partial\theta^{1}}-(\delta k_{4}^{2})^{*}\frac{\partial}{\partial\theta^{2}}-(\delta k_{4}^{3})^{*}\frac{\partial}{\partial\theta^{3}}+i\,\delta a_{4}\frac{\partial}{\partial\theta^{4}}-\delta k_{8}^{1}\theta^{1}-\delta k_{8}^{2}\theta^{2}-\delta k_{8}^{3}\theta^{3},
\end{array}
\end{equation}
а (25) переходит в (30) 
\begin{equation}
\begin{array}{c}
(\theta^{1})'=\theta^{1}-i\,\delta a_{1}\theta^{1}+(\delta k_{2}^{1})^{*}\theta^{2}+(\delta k_{3}^{1})^{*}\theta^{3}+(\delta k_{4}^{1})^{*}\theta^{4}-(\delta k_{6}^{1})^{*}\frac{\partial}{\partial\theta^{2}}-(\delta k_{7}^{1})^{*}\frac{\partial}{\partial\theta^{3}}-(\delta k_{8}^{1})^{*}\frac{\partial}{\partial\theta^{4}},\\
(\theta^{2})'=\theta^{2}-\delta k_{2}^{1}\theta^{1}-i\,\delta a_{2}\theta^{2}+(\delta k_{3}^{2})^{*}\theta^{3}+(\delta k_{4}^{2})^{*}\theta^{4}+(\delta k_{6}^{1})^{*}\frac{\partial}{\partial\theta^{1}}-(\delta k_{7}^{2})^{*}\frac{\partial}{\partial\theta^{3}}-(\delta k_{8}^{2})^{*}\frac{\partial}{\partial\theta^{4}},\\
(\theta^{3})'=\theta^{3}-\delta k_{3}^{1}\theta^{1}-\delta k_{3}^{2}\theta^{2}-i\,\delta a_{3}\theta^{3}+(\delta k_{4}^{3})^{*}\theta^{4}+(\delta k_{7}^{1})^{*}\frac{\partial}{\partial\theta^{1}}+(\delta k_{7}^{2})^{*}\frac{\partial}{\partial\theta^{2}}-(\delta k_{8}^{3})^{*}\frac{\partial}{\partial\theta^{4}},\\
(\theta^{4})'=\theta^{4}-\delta k_{4}^{1}\theta^{1}-\delta k_{4}^{2}\theta^{2}-\delta k_{4}^{3}\theta^{3}-i\,\delta a_{4}\theta^{4}+(\delta k_{8}^{1})^{*}\frac{\partial}{\partial\theta^{1}}+(\delta k_{8}^{2})^{*}\frac{\partial}{\partial\theta^{2}}+(\delta k_{8}^{3})^{*}\frac{\partial}{\partial\theta^{3}}.
\end{array}
\end{equation}

Легко убедиться, что преобразования (29) и (30) осуществляются инфинитезимальным
оператором 
\begin{equation}
e^{d\hat{G}}=1+d\hat{G},
\end{equation}
где 
\begin{equation}
\begin{array}{ccc}
d\hat{G} & = & d\hat{G}_{1}+d\hat{G}_{2}+d\hat{G}_{3}+d\hat{G}_{4}+d\hat{G}_{5}+d\hat{G}_{6}+d\hat{G}_{7},\\
d\hat{G}_{1} & = & \int d^{3}p\,i\,[\delta a_{1}\frac{\partial}{\partial\theta^{1}}\theta^{1}+\delta a_{2}\frac{\partial}{\partial\theta^{2}}\theta^{2}+\delta a_{3}\frac{\partial}{\partial\theta^{3}}\theta^{3}+\delta a_{4}\frac{\partial}{\partial\theta^{4}}\theta^{4},*],\\
d\hat{G}_{2} & = & \int d^{3}p\,[(\delta k_{8}^{1})^{*}\frac{\partial}{\partial\theta^{1}}\frac{\partial}{\partial\theta^{4}}+\delta k_{8}^{1}\theta^{4}\theta^{1}+(\delta k_{7}^{2})^{*}\frac{\partial}{\partial\theta^{2}}\frac{\partial}{\partial\theta^{3}}+\delta k_{7}^{2}\theta^{3}\theta^{2},*],\\
d\hat{G}_{3} & = & \int d^{3}p\,[(\delta k_{7}^{1})^{*}\frac{\partial}{\partial\theta^{1}}\frac{\partial}{\partial\theta^{3}}+\delta k_{7}^{1}\theta^{3}\theta^{1}+(\delta k_{8}^{2})^{*}\frac{\partial}{\partial\theta^{2}}\frac{\partial}{\partial\theta^{4}}+\delta k_{8}^{2}\theta^{4}\theta^{2},*],\\
d\hat{G}_{4} & = & \int d^{3}p\,[(\delta k_{6}^{1})^{*}\frac{\partial}{\partial\theta^{1}}\frac{\partial}{\partial\theta^{2}}+\delta k_{6}^{1}\theta^{2}\theta^{1}+(\delta k_{8}^{3})^{*}\frac{\partial}{\partial\theta^{3}}\frac{\partial}{\partial\theta^{4}}+\delta k_{8}^{3}\theta^{4}\theta^{3},*],\\
d\hat{G}_{5} & = & \int d^{3}p\,[\delta k_{2}^{1}\frac{\partial}{\partial\theta^{2}}\theta^{1}-(\delta k_{2}^{1})^{*}\frac{\partial}{\partial\theta^{1}}\theta^{2}+\delta k_{4}^{3}\frac{\partial}{\partial\theta^{4}}\theta^{3}-(\delta k_{4}^{3})^{*}\frac{\partial}{\partial\theta^{3}}\theta^{4},*],\\
d\hat{G}_{6} & = & \int d^{3}p\,[\delta k_{3}^{1}\frac{\partial}{\partial\theta^{3}}\theta^{1}-(\delta k_{3}^{1})^{*}\frac{\partial}{\partial\theta^{1}}\theta^{3}+\delta k_{4}^{2}\frac{\partial}{\partial\theta^{4}}\theta^{2}-(\delta k_{4}^{2})^{*}\frac{\partial}{\partial\theta^{2}}\theta^{4},*],\\
d\hat{G}_{7} & = & \int d^{3}p\,[\delta k_{4}^{1}\frac{\partial}{\partial\theta^{4}}\theta^{1}-(\delta k_{4}^{1})^{*}\frac{\partial}{\partial\theta^{1}}\theta^{4}+\delta k_{3}^{2}\frac{\partial}{\partial\theta^{3}}\theta^{2}-(\delta k_{3}^{2})^{*}\frac{\partial}{\partial\theta^{2}}\theta^{3},*].
\end{array}
\end{equation}

Введем величины

\begin{equation}
\begin{array}{ccc}
\hat{\gamma}^{0} & = & \int d^{3}p\,[\frac{\partial}{\partial\theta^{1}(p)}\theta^{1}(p)+\frac{\partial}{\partial\theta^{2}(p)}\theta^{2}(p)+\frac{\partial}{\partial\theta^{3}(p)}\theta^{3}(p)+\frac{\partial}{\partial\theta^{4}(p)}\theta^{4}(p),*],\\
\hat{\gamma}^{1} & = & \int d^{3}p\,[\frac{\partial}{\partial\theta^{2}(p)}\frac{\partial}{\partial\theta^{3}(p)}-\theta^{3}(p)\theta^{2}(p)+\frac{\partial}{\partial\theta^{1}(p)}\frac{\partial}{\partial\theta^{4}(p)}-\theta^{4}(p)\theta^{1}(p),*],\\
\hat{\gamma}^{2} & = & i\int d^{3}p\,[\frac{\partial}{\partial\theta^{2}(p)}\frac{\partial}{\partial\theta^{3}(p)}+\theta^{3}(p)\theta^{2}(p)-\frac{\partial}{\partial\theta^{1}(p)}\frac{\partial}{\partial\theta^{4}(p)}-\theta^{4}(p)\theta^{1}(p),*],\\
\hat{\gamma}^{3} & = & \int d^{3}p\,[\frac{\partial}{\partial\theta^{1}(p)}\frac{\partial}{\partial\theta^{3}(p)}-\theta^{3}(p)\theta^{1}(p)-\frac{\partial}{\partial\theta^{2}(p)}\frac{\partial}{\partial\theta^{4}(p)}+\theta^{4}(p)\theta^{2}(p),*],\\
\hat{\gamma}^{4} & = & i\int d^{3}p\,[\frac{\partial}{\partial\theta^{1}(p)}\frac{\partial}{\partial\theta^{3}(p)}+\theta^{3}(p)\theta^{1}(p)+\frac{\partial}{\partial\theta^{2}(p)}\frac{\partial}{\partial\theta^{4}(p)}+\theta^{4}(p)\theta^{2}(p),*],\\
\hat{\gamma}^{6} & = & i\int d^{3}p\,[\frac{\partial}{\partial\theta^{1}(p)}\frac{\partial}{\partial\theta^{2}(p)}+\theta^{2}(p)\theta^{1}(p)-\frac{\partial}{\partial\theta^{3}(p)}\frac{\partial}{\partial\theta^{4}(p)}-\theta^{4}(p)\theta^{3}(p),*],\\
\hat{\gamma}^{7} & = & \int d^{3}p\,[\frac{\partial}{\partial\theta^{1}(p)}\frac{\partial}{\partial\theta^{2}(p)}-\theta^{2}(p)\theta^{1}(p)+\frac{\partial}{\partial\theta^{3}(p)}\frac{\partial}{\partial\theta^{4}(p)}-\theta^{4}(p)\theta^{3}(p),*],
\end{array}
\end{equation}

Величины $\hat{\gamma}^{\mu},\,\mu=0,1,2,3$ из (33) являются супералгебраическим
аналогом соответствующих матриц Дирака $\gamma^{\mu}$, при этом $\hat{\gamma}^{4}$
соответствует $i\gamma^{5}$, а $\hat{\gamma}^{6},\hat{\gamma}^{7}$
\textendash{} базисные клиффордовы векторы, перемешивающие компоненты
спиноров и сопряженных спиноров, не имеющие аналогов в теории Драка.
При этом $(\hat{\gamma}^{4})^{2}=(\hat{\gamma}^{6})^{2}=(\hat{\gamma}^{7})^{2}=-1$.
Необходимо отметить, что из семи данных величин только шесть являются
независимыми, так как их произведение равно $-i$. Первые четыре соответствуют
осям пространства-времени, $\hat{\gamma}^{5}=-i\hat{\gamma}^{4}$
является псевдовектором, а $\hat{\gamma}^{6}$ и $\hat{\gamma}^{7}$
не имеют аналогов в матричной теории спиноров и задают двумерное пространство
внутренних степеней свободы спиноров.

Обозначим 
\[
\begin{array}{c}
dA_{i}=Re(\delta k_{i}^{1}),\,da_{i}=Im(\delta k_{i}^{1}),\,dB_{i}=Re(\delta k_{i}^{2}),\\
db_{i}=Im(\delta k_{i}^{2}),\,dC_{i}=Re(\delta k_{i}^{3}),\,dc_{i}=Im(\delta k_{i}^{3}).
\end{array}
\]

Тогда 
\[
\begin{array}{cc}
d\hat{G}_{1}= & i\int d^{3}p\,[\frac{da_{1}+da_{2}+da_{3}+da_{4}}{2}(\frac{\partial}{\partial\theta^{1}}\theta^{1}+\frac{\partial}{\partial\theta^{2}}\theta^{2}+\frac{\partial}{\partial\theta^{3}}\theta^{3}+\frac{\partial}{\partial\theta^{4}}\theta^{4})-\\
 & \frac{da_{1}-da_{2}-da_{3}+da_{4}}{2}(\frac{\partial}{\partial\theta^{1}}\theta^{1}-\frac{\partial}{\partial\theta^{2}}\theta^{2}-\frac{\partial}{\partial\theta^{3}}\theta^{3}+\frac{\partial}{\partial\theta^{4}}\theta^{4})+\\
 & \frac{da_{1}-da_{2}+da_{3}-da_{4}}{2}(\frac{\partial}{\partial\theta^{1}}\theta^{1}-\frac{\partial}{\partial\theta^{2}}\theta^{2}+\frac{\partial}{\partial\theta^{3}}\theta^{3}-\frac{\partial}{\partial\theta^{4}}\theta^{4})+\\
 & \frac{da_{1}+da_{2}-da_{3}-da_{4}}{2}(\frac{\partial}{\partial\theta^{1}}\theta^{1}+\frac{\partial}{\partial\theta^{2}}\theta^{2}-\frac{\partial}{\partial\theta^{3}}\theta^{3}-\frac{\partial}{\partial\theta^{4}}\theta^{4}),*].
\end{array}
\]

То есть 
\begin{equation}
\begin{array}{cc}
d\hat{G}_{1}= & i\frac{da_{1}+da_{2}+da_{3}+da_{4}}{2}\hat{\gamma}^{0}+\frac{da_{1}-da_{2}-da_{3}+da_{4}}{2}\hat{\gamma}^{1}\hat{\gamma}^{2}+\\
 & \frac{da_{1}-da_{2}+da_{3}-da_{4}}{2}\hat{\gamma}^{3}\hat{\gamma}^{4}+\frac{-da_{1}-da_{2}+da_{3}+da_{4}}{2}\hat{\gamma}^{6}\hat{\gamma}^{7}.
\end{array}
\end{equation}

Обозначим 
\[
\begin{array}{c}
d\omega_{0}=\frac{da_{1}+da_{2}+da_{3}+da_{4}}{2},\,d\omega_{12}=\frac{da_{1}-da_{2}-da_{3}+da_{4}}{2},\\
d\omega_{34}=\frac{da_{1}-da_{2}+da_{3}-da_{4}}{2},\,d\omega_{67}=\frac{-da_{1}-da_{2}+da_{3}+da_{4}}{2}.
\end{array}
\]

Тогда (34) преобразуется в 
\begin{equation}
d\hat{G}_{1}=i\hat{\gamma}^{0}d\omega_{0}+\hat{\gamma}^{1}\hat{\gamma}^{2}d\omega_{12}+\hat{\gamma}^{3}\hat{\gamma}^{4}d\omega_{34}+\hat{\gamma}^{6}\hat{\gamma}^{7}d\omega_{67}.
\end{equation}

Аналогично, обозначив 
\[
\begin{array}{c}
d\omega_{01}=\frac{dA_{8}+dB_{7}}{2},\,d\omega_{2}=\frac{dA_{8}-dB_{7}}{2},\,d\omega_{1}=\frac{da_{8}+db_{7}}{2},\,d\omega_{02}=\frac{-da_{8}+db_{7}}{2},\\
d\omega_{4}=-\frac{dA_{7}+dB_{8}}{2},\,d\omega_{03}=\frac{dA_{7}-dB_{8}}{2},\,d\omega_{04}=\frac{da_{7}+db_{8}}{2},\,d\omega_{3}=\frac{da_{7}-db_{8}}{2},\\
d\omega_{07}=\frac{dA_{6}+dC_{8}}{2},\,d\omega_{6}=-\frac{dA_{6}-dC_{8}}{2},\,d\omega_{7}=\frac{da_{6}+dc_{8}}{2},\,d\omega_{06}=\frac{da_{6}-dc_{8}}{2},\\
d\omega_{13}=-\frac{dA_{2}+dC_{4}}{2},\,d\omega_{24}=-\frac{dA_{2}-dC_{4}}{2},\,d\omega_{14}=-\frac{da_{2}+dc_{4}}{2},\,d\omega_{23}=\frac{da_{2}-dc_{4}}{2},\\
d\omega_{26}=\frac{dA_{3}+dB_{4}}{2},\,d\omega_{17}=-\frac{dA_{3}-dB_{4}}{2},\,d\omega_{27}=-\frac{da_{3}+db_{4}}{2},\,d\omega_{16}=-\frac{da_{3}-db_{4}}{2},\\
d\omega_{37}=-\frac{dA_{4}+dB_{3}}{2},\,d\omega_{46}=\frac{dA_{4}-dB_{3}}{2},\,d\omega_{36}=-\frac{da_{4}+db_{3}}{2},\,d\omega_{47}=\frac{da_{4}-db_{3}}{2},
\end{array}
\]
получаем 
\begin{equation}
\begin{array}{c}
d\hat{G}_{2}=\hat{\gamma}^{0}\hat{\gamma}^{1}d\omega_{01}+i\hat{\gamma}^{2}d\omega_{2}+i\hat{\gamma}^{1}d\omega_{1}+\hat{\gamma}^{0}\hat{\gamma}^{2}d\omega_{02},\\
d\hat{G}_{3}=i\hat{\gamma}^{4}d\omega_{4}+\hat{\gamma}^{0}\hat{\gamma}^{3}d\omega_{03}+\hat{\gamma}^{0}\hat{\gamma}^{4}d\omega_{04}+i\hat{\gamma}^{3}d\omega_{3},\\
d\hat{G}_{4}=\hat{\gamma}^{0}\hat{\gamma}^{7}d\omega_{07}+i\hat{\gamma}^{6}d\omega_{6}+i\hat{\gamma}^{7}d\omega_{7}+\hat{\gamma}^{0}\hat{\gamma}^{6}d\omega_{06},\\
d\hat{G}_{5}=\hat{\gamma}^{1}\hat{\gamma}^{3}d\omega_{03}+\hat{\gamma}^{2}\hat{\gamma}^{4}d\omega_{24}+\hat{\gamma}^{1}\hat{\gamma}^{4}d\omega_{14}+\hat{\gamma}^{2}\hat{\gamma}^{3}d\omega_{23},\\
d\hat{G}_{6}=\hat{\gamma}^{2}\hat{\gamma}^{6}d\omega_{26}+\hat{\gamma}^{1}\hat{\gamma}^{7}d\omega_{17}+\hat{\gamma}^{2}\hat{\gamma}^{7}d\omega_{27}+\hat{\gamma}^{1}\hat{\gamma}^{6}d\omega_{16},\\
d\hat{G}_{7}=\hat{\gamma}^{3}\hat{\gamma}^{7}d\omega_{37}+\hat{\gamma}^{4}\hat{\gamma}^{6}d\omega_{46}+\hat{\gamma}^{3}\hat{\gamma}^{6}d\omega_{36}+\hat{\gamma}^{4}\hat{\gamma}^{7}d\omega_{47}.
\end{array}
\end{equation}

Из (31), (32), (35), (36) следует, что оператор поля, состоящий из
линейной комбинации величин $\partial/\partial\theta^{i}(p)$ и $\theta^{i}(p)$,
преобразуется как 
\begin{equation}
\Psi'=\Psi+i\hat{\gamma}^{j}d\omega_{j}\Psi+\hat{\gamma}^{k}\hat{\gamma}^{l}d\omega_{kl}\Psi,
\end{equation}
где $j,k,l=0,1,2,3,4,6,7,\,k<l$. Интегрирование (37) по инфинитезимальным
вещественным параметрам $d\omega_{j}$ и $d\omega_{kl}$ дает экспоненциальную
форму разложения с конечными параметрами: 
\begin{equation}
\Psi'=exp(\hat{\gamma}^{k}\hat{\gamma}^{l}\omega_{kl})\,exp(i\hat{\gamma}^{j}\omega_{j})\Psi.
\end{equation}

\section{Разложение оператора поля по импульсам для пространства с одной времениподобной
осью}

Разложение, аналогичное (38), известно для решений уравнения Дирака
в варианте Дирака-Кэлера, оно описывает дираковскую частицу с внутренними
степенями свободы {[}23{]}. Наличие в (38) дополнительных базисных
клиффордовых векторов по сравнению с обычной теорией Дирака увеличивает
число внутренних степеней свободы.

Смысл множителя $exp(\hat{\gamma}^{k}\hat{\gamma}^{l}\omega_{kl})$
очевиден \textendash{} это клиффордовы (в том числе лоренцевы) вращения.

Рассмотрим действие множителя $exp(i\hat{\gamma}^{j}\omega_{j})$
в системе покоя спинора, когда 
\[
\Psi=\phi^{\alpha}\frac{\partial}{\partial\theta^{\alpha}(0)}+\chi_{\tau}\theta^{\tau}(0),
\]
где $\alpha,\tau=1,2,3,4$, $\phi^{\alpha}$ и $\chi_{\tau}$ числовые
коэффициенты, и когда $\omega_{j}=0$ при $j\neq0$, то есть при $exp(i\hat{\gamma}^{j}\omega_{j})=exp(i\hat{\gamma_{0}}\omega^{0})$.
Обозначим $\omega^{0}=-x^{0}m$ , где $m$ \textendash{} масса спинора
(положительная вещественная константа), $p_{0}=m$ и $x^{0}$- компонента
некоторого 4-вектора. Тогда 
\[
\begin{array}{cc}
\Psi'= & exp(-i\hat{\gamma}_{0}mx^{0})(\phi^{\alpha}\frac{\partial}{\partial\theta^{\alpha}(0)}+\chi_{\tau}\theta^{\tau}(0))=\\
 & exp(-ip_{0}x^{0})\phi^{\alpha}\frac{\partial}{\partial\theta^{\alpha}(0)}+exp(ip_{0}x^{0})\chi_{\tau}\theta^{\tau}(0).
\end{array}
\]

При преобразовании Лоренца $exp(\hat{\gamma}^{k}\hat{\gamma}^{l}\omega_{kl})$
величина $exp(-i\hat{\gamma}_{0}mx^{0})$ перейдет в $exp(i\hat{\gamma}^{j}\omega_{j})$,
и с учетом того, что $p_{0}x^{0}=p''_{\mu}x''^{\mu}$, мы получим
\[
\Psi''=exp(\hat{\gamma}^{k}\hat{\gamma}^{l}\omega_{kl})\Psi'=\phi^{\alpha}\,exp(-ip''_{\mu}x''^{\mu})b_{\alpha}(p'')+\chi_{\tau}\,exp(ip''_{\mu}x''^{\mu})\bar{b}_{\tau}(p''),
\]
где $b_{\alpha}(p)$ и $\bar{b}_{\tau}(p)$ \textendash{} операторы
из (24), (25). Как уже отмечалось, они обладают такими же антикоммутационными
свойствами, как обычные операторы рождения-уничтожения.

Интегрируя по всем возможным значениям импульса, получаем супералгебраическую
форму разложения вторичного квантования решений уравнения Дирака {[}16{]}:
\begin{equation}
\Psi(x)=\int d^{3}p\,\left(\phi^{\alpha}\,exp(-ip{}_{\mu}x{}^{\mu})b_{\alpha}(p)+\chi_{\tau}\,exp(ip{}_{\mu}x{}^{\mu})\bar{b}_{\tau}(p)\right)
\end{equation}

Откуда сразу следует физический смысл величин $x{}^{\mu}$ \textendash{}
это координаты в пространстве-времени.

Необходимо отметить, что разложение (39) по импульсам справедливо
только для времениподобного вектора $\hat{\gamma}^{j}\omega_{j}$,
в противном случае не существует преобразования Лоренца, переводящего
в него вектор $-\hat{\gamma}_{0}mx^{0}$. Вещественность массы спинора
обеспечивается вещественностью параметров $\omega_{j}$ в (38).

Если вектор $\hat{\gamma}^{j}\omega_{j}$ пространственноподобный,
системы покоя спинора не существует, но в качестве начального состояния
можно выбрать произвольную систему с равной нулю составляющей импульса
вдоль оси времени. В силу произвольности направления пространственных
осей выберем в качестве такой оси $\hat{\gamma}^{1}$, и будем рассматривать
инфинитезимальные преобразования. Обозначим $d\omega^{1}=-dx^{1}m$,
где, как и ранее, $m>0$. Тогда $\hat{\gamma}^{j}d\omega_{j}=-\hat{\gamma}^{1}m\,dx_{1}=-\hat{\gamma}_{1}m\,dx^{1}$,
$p=p_{1}=m$, и мы получаем 
\begin{equation}
\Psi=\phi^{\alpha}\frac{\partial}{\partial\theta^{\alpha}(p_{1})}+\chi_{\tau}\theta^{\tau}(p_{1}),
\end{equation}
\begin{equation}
\Psi'=(1-i\hat{\gamma}_{1}m\,dx^{1})\Psi.
\end{equation}

В (41) величина $dx^{1}$ бесконечно малая, но последовательное применение
преобразований (41) даст формулу для конечных значений $x^{1}$: 
\begin{equation}
\Psi'=exp(-i\hat{\gamma}_{1}mx^{1})\Psi.
\end{equation}

Перепишем разложение (40) в виде разложения по собственным векторам
$\hat{\gamma}_{1}$, опуская для краткости аргумент $p_{1}$: 
\[
\begin{array}{c}
\Psi=\Psi_{+}+\Psi_{-},\\
\hat{\gamma}_{1}\Psi_{+}=i\Psi_{+},\\
\hat{\gamma}_{1}\Psi_{-}=-i\Psi_{-},\\
\Psi_{+}=\psi_{1+}(\frac{\partial}{\partial\theta^{1}}+i\theta^{4})+\psi_{2+}(\frac{\partial}{\partial\theta^{2}}+i\theta^{3})+\psi_{3+}(\frac{\partial}{\partial\theta^{4}}-i\theta^{1})+\psi_{4+}(\frac{\partial}{\partial\theta^{3}}-i\theta^{2}),\\
\Psi_{-}=\psi_{1-}(\frac{\partial}{\partial\theta^{1}}-i\theta^{4})+\psi_{2-}(\frac{\partial}{\partial\theta^{2}}-i\theta^{3})+\psi_{3-}(\frac{\partial}{\partial\theta^{4}}+i\theta^{1})+\psi_{4-}(\frac{\partial}{\partial\theta^{3}}+i\theta^{2}),
\end{array}
\]
где $\psi_{1+}=(\phi^{1}-i\chi_{4})/2,\,\psi_{1-}=(\phi^{1}+i\chi_{4})/2$,
и так далее.

Тогда (42) можно переписать в виде 
\begin{equation}
\Psi'=exp(-iimx^{1})\Psi_{+}+exp(iimx^{1})\Psi_{-}=exp(mx^{1})\Psi_{+}+exp(-mx^{1})\Psi_{-}.
\end{equation}

Оба слагаемых в (43) являются расходящимися, первое при $x^{1}\rightarrow+\infty$,
а второе при $x^{1}\rightarrow-\infty$. Поэтому нормированное решение
(43) в любой конечной области оказывается равным нулю. Таким образом,
в супералгебраической теории спиноров тахионные решения имеют нулевую
амплитуду, то есть отсутствуют.

Если вектор $\hat{\gamma}^{j}\omega_{j}$ светоподобный, системы покоя
спинора не существует, но в качестве начального состояния можно выбрать
систему с равными составляющими импульса вдоль оси времени и вдоль
произвольной пространственной оси. Выберем в качестве такой оси, например,$\hat{\gamma}^{1}$
, при этом $\hat{\gamma}^{j}\omega_{j}=-(\hat{\gamma}^{0}+\hat{\gamma}^{1})\,mx_{1},\,p_{0}=p_{1}$
и $(\hat{\gamma}^{0}+\hat{\gamma}^{1})^{2}=0$.

Поскольку $\hat{\gamma}^{0}+\hat{\gamma}^{1}=\hat{\gamma}^{0}(1-\hat{\gamma}^{1}\hat{\gamma}^{0})$,
из-за чего $(\hat{\gamma}^{0}+\hat{\gamma}^{1})\frac{1+\hat{\gamma}^{1}\hat{\gamma}^{0}}{2}=0$
и $(\hat{\gamma}^{0}+\hat{\gamma}^{1})\frac{1-\hat{\gamma}^{1}\hat{\gamma}^{0}}{2}=\hat{\gamma}^{0}+\hat{\gamma}^{1}$
, разложим $\Psi$ по собственным векторам величины $\hat{\gamma}^{1}\hat{\gamma}^{0}$.
Ими являются величины $\frac{\partial}{\partial\theta^{1}}-\theta^{4},\,\frac{\partial}{\partial\theta^{2}}-\theta^{3},\,\frac{\partial}{\partial\theta^{4}}+\theta^{1},\,\frac{\partial}{\partial\theta^{3}}+\theta^{2}$
с собственным числом +1, и величины $\frac{\partial}{\partial\theta^{1}}+\theta^{4},\,\frac{\partial}{\partial\theta^{2}}+\theta^{3},\,\frac{\partial}{\partial\theta^{4}}-\theta^{1},\,\frac{\partial}{\partial\theta^{3}}-\theta^{2}$
с собственным числом -1. Тогда 
\[
\begin{array}{c}
\Psi=\Psi_{+}+\Psi_{-},\\
(\hat{\gamma}^{0}+\hat{\gamma}^{1})\Psi_{+}=0,\\
(\hat{\gamma}^{0}+\hat{\gamma}^{1})\Psi_{-}\neq0,\\
\Psi_{+}=\psi_{1+}(\frac{\partial}{\partial\theta^{1}}-\theta^{4})+\psi_{2+}(\frac{\partial}{\partial\theta^{2}}-\theta^{3})+\psi_{3+}(\frac{\partial}{\partial\theta^{4}}+\theta^{1})+\psi_{4+}(\frac{\partial}{\partial\theta^{3}}+\theta^{2}),\\
\Psi_{-}=\psi_{1-}(\frac{\partial}{\partial\theta^{1}}+\theta^{4})+\psi_{2-}(\frac{\partial}{\partial\theta^{2}}+\theta^{3})+\psi_{3-}(\frac{\partial}{\partial\theta^{4}}-\theta^{1})+\psi_{4-}(\frac{\partial}{\partial\theta^{3}}-\theta^{2}).
\end{array}
\]
Аналогом (37) без лоренцевых поворотов в этом случае будет 
\begin{equation}
\Psi'=(1-i(\hat{\gamma}^{0}+\hat{\gamma}^{1})p_{0}x_{1})(\Psi_{+}+\Psi_{-})=\Psi_{+}+\Psi_{-}-i(\hat{\gamma}^{0}+\hat{\gamma}^{1})p_{0}x_{1}\Psi_{-}.
\end{equation}

Из-за нильпотентности величины $\hat{\gamma}^{0}+\hat{\gamma}^{1}$
предыдущий подход, основанный на интегрировании инфинитезимальных
преобразований (37) и получении экспоненциальной формы (38) преобразований,
некорректен. Однако легко убедиться, что в этом случае справедливо
уравнение (44) для произвольных конечных значений $x_{1}$ \textendash{}
такое преобразование сохраняет коммутационные свойства $\partial'/\partial\theta^{i}$
и $(\theta^{j})'$.

Требование конечности $\Psi'$ приводит к равенству нулю последнего
слагаемого в (44), что выполняется только при $\Psi_{-}=0$. Поэтому
$\Psi=\Psi_{+}$ и $\Psi'=\Psi$. Таким образом, отсутствует зависимость
$\Psi'$ от координат и времени, и разложение по импульсам для спиноров
с нулевой массой в рамках супералгебраической теории спиноров не возникает.

В итоге можно констатировать, что в супералгебраической теории оказываются
возможны только спиноры с времениподобными импульсами, и эта особенность
обусловлена чисто алгебраическими причинами, без привлечения принципа
причинности.

Совершенно аналогично производится разложение по импульсам в случае
других сигнатур пространства-времени. При этом сигнатура базисных
клиффордовых векторов, получающихся в формулах, аналогичных (35)-(36),
определяется вариантом дираковского сопряжения (22).

\section{Оператор заряда для пространства с одной времениподобной осью}

Легко видеть, что оператор поля (39) удовлетворяет уравнению Дирака:
\begin{equation}
\hat{\gamma}^{\mu}i\frac{\partial}{\partial x^{\mu}}\Psi(x)=m\Psi(x).
\end{equation}

В то же время этот оператор содержит в два раза больше компонент,
чем обычный 4-спинор Дирака, так как в нем присутствуют как компоненты
спинора, так и компоненты сопряженного спинора. Ведь в супералгебраическом
представлении они удовлетворяют одному и тому же уравнению Дирака
{[}16{]}. Для решения этой проблемы рассмотрим набор взаимно коммутирующих
линейно независимых операторов, построенных из базисных клиффордовых
векторов. В таком качестве можно взять плотности в импульсном пространстве
при $p=0$ 
\[
\begin{array}{ccc}
\hat{\gamma}^{0}(0) & = & [\frac{\partial}{\partial\theta^{1}}\theta^{1}+\frac{\partial}{\partial\theta^{2}}\theta^{2}+\frac{\partial}{\partial\theta^{3}}\theta^{3}+\frac{\partial}{\partial\theta^{4}}\theta^{4},*],\\
\hat{S}_{3}(0) & = & \frac{i}{2}(\hat{\gamma}^{1}\hat{\gamma}^{2})(0)[\frac{\partial}{\partial\theta^{1}}\theta^{1}-\frac{\partial}{\partial\theta^{2}}\theta^{2}-\frac{\partial}{\partial\theta^{3}}\theta^{3}+\frac{\partial}{\partial\theta^{4}}\theta^{4},*],\\
\hat{W}_{3}(0)/m & = & -\frac{i}{2}(\hat{\gamma}^{3}\hat{\gamma}^{4})(0)[\frac{\partial}{\partial\theta^{1}}\theta^{1}-\frac{\partial}{\partial\theta^{2}}\theta^{2}+\frac{\partial}{\partial\theta^{3}}\theta^{3}-\frac{\partial}{\partial\theta^{4}}\theta^{4},*],\\
\hat{Q}(0) & = & -i(\hat{\gamma}^{6}\hat{\gamma}^{7})(0)=[-\frac{\partial}{\partial\theta^{1}}\theta^{1}-\frac{\partial}{\partial\theta^{2}}\theta^{2}+\frac{\partial}{\partial\theta^{3}}\theta^{3}+\frac{\partial}{\partial\theta^{4}}\theta^{4},*],
\end{array}
\]
где $\hat{S}_{3}$ и $\hat{W}_{3}$ \textendash{} компоненты операторов
спина и связанной со спином части вектора Паули-Любанского, а $\hat{Q}(p)$
\textendash{} оператор плотности некоторого заряда. Интегрируя по
всем импульсам для различных лоренцевых поворотов, приходим к формуле
заряда 
\begin{equation}
\hat{Q}=\int d^{3}p\,[\bar{b}_{1}(p)b_{1}(p)+\bar{b}_{2}(p)b_{2}(p)-\bar{b}_{3}(p)b_{3}(p)-\bar{b}_{4}(p)b_{4}(p),*].
\end{equation}

Рассмотрим собственные векторы $\Psi$ и собственные числа $q$ оператора
$\hat{Q}$. При действии на вакуум $\hat{Q}\Psi_{V}=0$. Для спиноров
из (46) и (39) следует, что $\hat{Q}\Psi=q\Psi$ только если четыре
компоненты восьмикомпонентного спинора $\Psi$ нулевые \textendash{}
а ненулевые, например, при $\alpha=1,2,\,\tau=3,4$. В этом случае
$\hat{Q}\Psi=\Psi$, $\hat{Q}\bar{\Psi}=-\bar{\Psi}$. 

Таким образом, если потребовать, чтобы спиноры являлись собственными
векторами оператора заряда, количество компонент спиноров и сопряженных
спиноров уменьшается до четырех, как в теории Дирака. При этом спиноры
и сопряженные спиноры обладают противоположными значениями данного
заряда, а вакуум обладает нулевым зарядом.

Существование подобного заряда спиноров с аналогичной $\hat{Q}$ формулой
хорошо известно в теории вторичного квантования, и обычно его отождествляют
с электрическим зарядом {[}24{]}. Обычным путем {[}24{]} легко показать,
что из уравнения (45) следует, что этот заряд является интегралом
движения.

\section{Заключение}

Таким образом, построена супералгебраическая теория спиноров, которая
автоматически является вторично квантованной. В этой теории введено
скалярное произведение для операторов поля и векторов состояний, и
для пространств произвольной размерности и сигнатур введено обобщение
дираковского сопряжения, обеспечивающее Лоренц-ковариантное преобразование
сопряженных спиноров.

Показано, что сигнатура базисных клиффордовых векторов, являющихся
аналогом гамма-матриц Дирака, и их количество, а также разложение
вторичного квантования по импульсам задаются вариантом обобщенного
дираковского сопряжения и требованием сохранения при преобразованиях
спиноров и сопряженных спиноров их CAR-алгебры. При этом разложение
вторичного квантования по импульсам оказывается возможным только для
спиноров с времениподобными импульсами, то есть с ненулевой массой
и не нарушающих причинность. Такое разложение имеет чисто алгебраическую
природу, и для его построения нет необходимости привлекать принцип
причинности.

В матричной теории дираковские спиноры и сопряженные к ним спиноры
существуют в разных пространствах (сопряженных). Однако в теории вторичного
квантования операторы поля должны существовать в одном и том же пространстве,
что и реализуется в построенной супералгебраической теории спиноров.
Благодаря этому в ней возникают два дополнительных базисных клиффордовых
вектора, порождающие внутреннюю степень свободы спиноров. Оператор
вращения в соответствующей этим векторам плоскости является оператором
заряда, аналогичный оператору электрического заряда в обычной теории
вторичного квантования спиноров.

\end{document}